\documentclass{emulateapj}
\usepackage{verbatim}
\usepackage{color}
\usepackage{amsmath}
\usepackage{graphicx}
\usepackage[flushleft]{threeparttable}

\usepackage{longtable}

\newcommand{\uvflux}{8.4 $\times 10^{-15}$ erg s$^{-1}$ cm$^{-2}$}
\newcommand{\xuvfrac}{3.4 $\times 10^{-4} L_{bol}$}

\bibpunct[ ]{(}{)}{;}{}{}{,}
\shorttitle{TRAPPIST-1}
\shortauthors{Becker et al.}

\begin{document}
\title{A Coupled Analysis of Atmospheric Mass Loss and Tidal Evolution in XUV Irradiated Exoplanets: the TRAPPIST-1 Case Study}
\author{Juliette Becker\altaffilmark{1,2,3, *}, Elena Gallo\altaffilmark{1}, Edmund Hodges-Kluck\altaffilmark{4}, Fred C. Adams\altaffilmark{1,3}, Rory Barnes\altaffilmark{5,6}}
\affil{\textsuperscript{1}Department of Astronomy, University of Michigan, Ann Arbor, MI 48104, USA \\}
\affil{\textsuperscript{2}Division of Geological and Planetary Sciences, California Institute of Technology, Pasadena, CA 91125 \\}
\affil{\textsuperscript{3}Department of Physics, University of Michigan, Ann Arbor, MI 48104, USA \\}
\affil{\textsuperscript{4}Code 662, NASA Goddard Space Flight Center, Greenbelt, MD 20771, USA\\}
\affil{\textsuperscript{5}Department of Astronomy, University of Washington, Seattle, WA, USA \\}
\affil{\textsuperscript{6}NASA Virtual Planetary Laboratory, USA \\}
\affil{\textsuperscript{*}51 Pegasi b Fellow\\}

\begin{abstract}
Exoplanets residing close to their stars can experience evolution of both their physical structures and their orbits due to the influence of their host stars. 
In this work, we present a coupled analysis of dynamical tidal dissipation and atmospheric mass loss for exoplanets in XUV irradiated environments.
As our primary application, we use this model to study the TRAPPIST-1 system, and place constraints on the interior structure and orbital evolution of the planets.
We start by reporting on a UV continuum flux measurement {(centered around $\sim1900$ Angstroms) for the star TRAPPIST-1}, based on 300 ks of {Neil Gehrels Swift Observatory} data, and which {enables an estimate of} the XUV-driven thermal escape arising from XUV photo-dissociation for each planet. {We find that the X-ray flaring luminosity, measured from our X-ray detections, of TRAPPIST-1 is 5.6 $\times$10$^{-4} L_{*}$, while the full flux including non-flaring periods is 6.1 $\times$10$^{-5} L_{*}$, when $L_{*}$ is TRAPPIST-1's bolometric luminosity.}
We then construct a model that includes both atmospheric mass-loss and tidal evolution, and requires the planets to attain their present-day orbital elements during this coupled evolution. We use this model to constrain the ratio $Q'=3Q/2k_{2}$ for each planet. Finally, we use additional numerical models implemented with the Virtual Planet Simulator \texttt{VPLanet} to study ocean retention for these planets using our derived system parameters.
\end{abstract}



\section{Introduction}\label{sec:intro}

TRAPPIST-1 provides a rare opportunity to study multi-planet systems around stars at the low-mass edge of the main sequence. The star TRAPPIST-1 is a M8V dwarf that hosts seven transiting planets in a highly compact configuration. 
The discovery of the first three planets was reported in \citet{Gillon2016}, followed by the subsequent detection of four more planets \citep{Gillon2017}. Since TRAPPIST-1 has a V-band magnitude of 18.8 \citep{Winters2015}, mass determination for the planets is difficult with radial velocity methods. Instead, mass estimates and constraints on other orbital parameters have been constructed using transit timing variations (TTVs) \citep[e.g.,][]{Gillon2017, Grimm2018} 
One major goal in the dynamical studies of planetary bodies is to understand their interior and exterior structures, which provide insights regarding how they formed. Planetary interior structures are often characterized in terms of a concentration parameter $k_2$ (the Love number) and their tidal dissipation susceptibilities $Q$ (the tidal quality factor). The exterior
structures of the planets, in contrast, are characterized by the chemical abundances in their atmospheres, which are influenced by atmospheric mass loss rates. 

In recent years, the increasing quantity of observational data has allowed some constraints to be placed on the $k_{2}$ and $Q$ values of exoplanets \citep[e.g.,][]{Jackson2008, Batygin2009,Kramm2012,Becker2013,Buhler2016,Csizmadia2019}. 

Simultaneously, as the planets evolve dynamically, their atmospheres also experience mass loss over time due to the XUV irradiation provided by the host star. 
The XUV radiation that allows atmospheric mass loss includes both X-ray photons (with wavelengths between $\sim$1 and 100 Angstroms) and EUV photons (with wavelengths between 100 and 912 Angstroms). The specific mechanism that drives mass loss, and the structure of flows driven by the XUV flux, depends on the planetary properties and stellar luminosity levels \citep{Owen2012}. Nonetheless, the general process involves the EUV radiation ionizing hydrogen atoms and heating the gas in the upper planetary atmosphere, and the X-ray radiation penetrating farther and driving an outward flow from deeper in the atmosphere. The net effect of both of these types of radiation flux is to break down the more massive molecules into ionized forms in the upper atmosphere; these components are liberated more easily and lead to atmospheric mass loss {\citep{VidalMadjar2003, Lammer2003}}. 

In this work, we model the planetary and stellar tidal dissipation and planetary atmospheric mass loss for exoplanets experiencing XUV irradiation. We then use this model to study the TRAPPIST-1 system, and place constraints on the interior structure and orbital evolution of the planets. Specifically, we constrain the allowed values of the $Q'$ ratio for each of the planets, with the four inner planets producing independent lower limits on their planetary $Q'$ ratios.


This paper is organized as follows. In Section \ref{sec:data}, we describe the \textit{Neil Gehrels Swift Observatory} (\textit{Swift}) data set and derive fundamental parameters for the host star TRAPPIST-1 and its planets. In Section \ref{sec:analysis_secular}, we discuss the theoretical insights that can be gained by considering a simultaneous analysis of both mass loss and planetary migration in these systems. Section \ref{sec:vplanet} considers how the measured and estimated parameters (the XUV flux and planetary $Q'$ ratios) affect ocean retention for each planet in the system. In Section \ref{sec:discuss}, we consider the implications of our results on the radiation environments and likely dynamical properties of the TRAPPIST-1 planets. Finally, the paper ends in Section \ref{sec:sum} with a summary of our results and suggestions for future lines of inquiry.

\section{Data and Derived Parameters}\label{sec:data}
{\subsection{SWIFT Data}}
Our data set consists of 34 epochs of observations of TRAPPIST-1 using the \textit{Swift} Ultraviolet/Optical Telescope (UVOT), for a total of $\sim$300 ks on target. The observations were taken between September 14, 2017 and July 8, 2018 using the \texttt{uvw2} filter, which has a central wavelength of 1928 Angstroms and a full width at half maximum of 657 Angstroms \citep{Poole2008}. {TRAPPIST-1 was also observed with XRT in photon-counting mode simultaneously with the UVOT measurements, and recorded photons with energies between 0.2 and 10 keV \citep{Burrows2005}. These two data sources allow a characterization of the UV and X-ray behavior of TRAPPIST-1 in both flaring and quiescent states. We describe the ways in which the data was processed in this section.}

\subsubsection{SWIFT UVOT Dataset}

This filter provides a measurement of the UV continuum to the redder side of the Lyman-$\alpha$ feature, which is generally used as a proxy of the continuum flux in the UV {\citep{Bourrier2017a, Bourrier2017}.}

The UVOT data were obtained in event (photon-counting) mode and reprocessed using the standard pipeline. We then used the XSELECT program\footnote{https://heasarc.gsfc.nasa.gov/ftools/xselect/} to remove periods of background flaring (defined as departures 3$\sigma$ above the mean {for each exposure; this is typically minimal for SWIFT and resulted in very few removals}) and to measure the flux through aperture photometry in each observation. 

\begin{figure*} 
   \centering
   \includegraphics[width=6.8in]{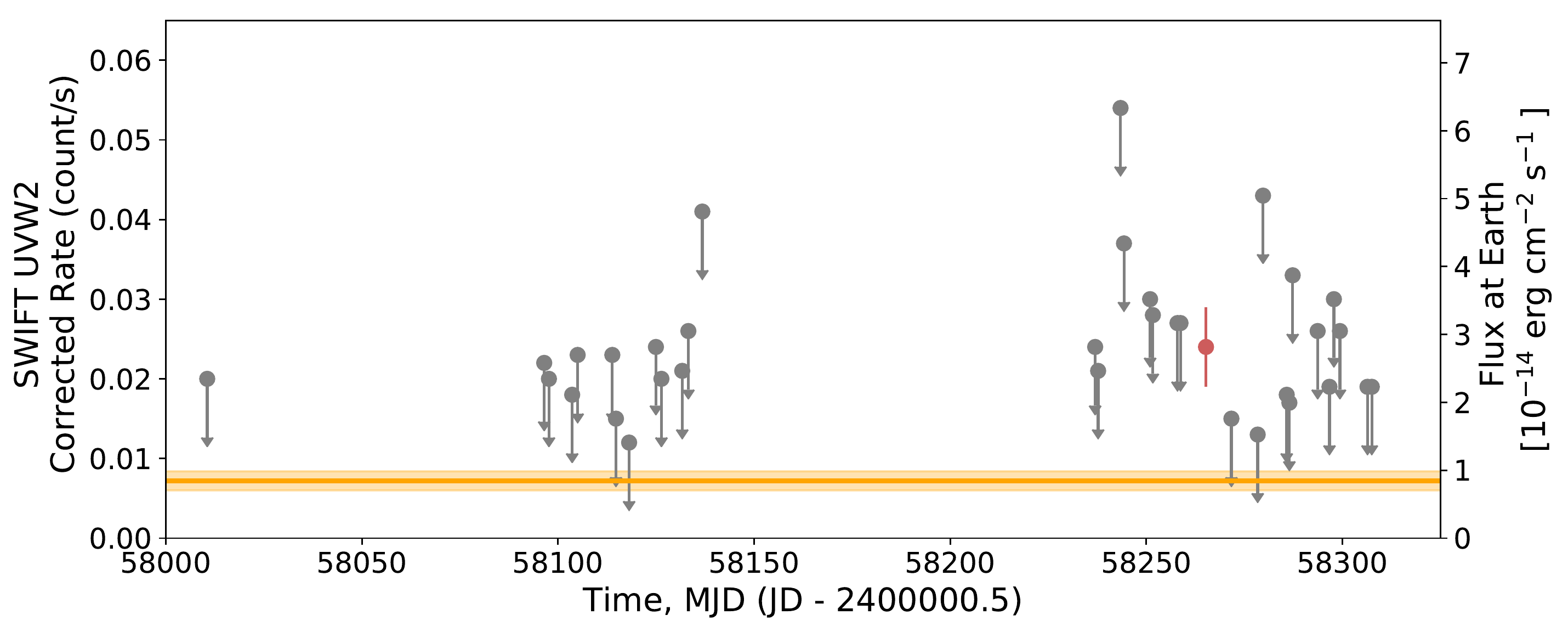}
   \caption{The UV light curve (with a waveband centered at 1928 Angstroms and a full width at half max of 657 Angstroms) constructed using SWIFT UVOT observations taken with the \texttt{uvw2} filter. {The single detection is plotted as a red point with error bars}, and upper limits are plotted as grey points with error arrows extending downwards. The average flux over the full 300ks of observations is plotted as an orange line, with the derived $1\sigma$ uncertainty plotted as a filled orange background region. {We note that the average flux was derived from the stacked 300ks image, rather than from the points plotted here.} The data used to plot this lightcurve is also given in Table \ref{tab:swiftlc}. } \label{fig:lightcurve}
\end{figure*}

Source regions are defined using Source Extractor \citep{bertin1996} using identical extraction radii (5~arcseconds) and background regions for each exposures. {The background region is used as a local background to estimate the background level in the detection cell.} The aperture photometry measurements are presented in Table~\ref{tab:swiftlc}. We note that this procedure, which gives us more control over the good time intervals, is consistent with aperture photometry on the standard SWIFT image products. The light curve from these measurements is shown in Figure~\ref{fig:lightcurve}.

We note that the SWIFT {\tt uvw2} filter has a known red transmission leak (non-zero transmission in the optical band). For red sources, optical light can significantly contaminate the {\tt uvw2} band. We therefore estimated the contribution from the Johnson $U$ and $V$ band fluxes for TRAPPIST-1 (assuming a quiescent, average value over the SWIFT baseline), as the red leak is negligible beyond $V$. We find that, at most, the contribution is about 0.001 counts per second, which is consistent with the statistical uncertainty.

\begin{table}
\begin{center}
\begin{tabular}{cccc}
\textbf{Obs. ID} & \textbf{Observation}  &\textbf{Observation}  & \textbf{Counts / sec} \\
 & \textbf{Start }  &\textbf{End }  &  \\
 & (JD-2400000.5) & (JD-2400000.5) & \\
\hline
10283001 &   58010.35 &  58010.77 & $<$ 0.020 \\ 
10283002 &   58095.98 &  58096.98 & $<$ 0.022 \\ 
10283003 &   58097.65 &  58097.72 &  $<$ 0.020 \\ 
10283004 &   58103.41 &  58103.81 &  $<$ 0.018 \\ 
10283005 &   58105.00 &  58105.01 &  $<$ 0.023 \\ 
10283006 &   58113.71 &  58113.99 &  $<$ 0.023 \\ 
10283007 &   58114.70 &  58114.85 &  $<$ 0.015 \\ 
10283008 &   58117.96 &  58118.31 &  $<$ 0.012 \\ 
10283009 &   58124.93 &  58125.00 &  $<$ 0.024 \\ 
10283010 &   58126.32 &  58126.41 &  $<$ 0.020 \\ 
10283011 &   58131.50 &  58131.84 &  $<$ 0.021 \\ 
10283012 &   58133.17 &  58133.30 &  $<$ 0.026 \\ 
10283013 &   58136.64 &  58136.96 &  $<$ 0.041 \\ 
10283014 &   58236.92 &  58237.00 &  $<$ 0.024 \\ 
10283015 &   58237.51 &  58237.92 &  $<$ 0.021 \\ 
10283016 &   58243.30 &  58243.56 &  $<$ 0.054 \\ 
10283017 &   58244.30 &  58244.31 &  $<$ 0.037 \\ 
10283018 &   58250.93 &  58251.00 &  $<$ 0.030 \\ 
10283019 &   58251.52 &  58251.81 &  $<$ 0.028 \\ 
10283020 &   58257.90 &  58257.99 &  $<$ 0.027 \\ 
10283021 &   58258.71 &  58258.72 &  $<$ 0.027 \\ 
10283022 &   58264.81 &  58265.63 &  0.024 $\pm$ 0.005 \\ 
10283023 &   58271.52 &  58271.87 &  $<$ 0.015 \\ 
10283024 &   58278.02 &  58278.77 &  $<$ 0.013 \\ 
10283025 &   58279.75 &  58279.77 &  $<$ 0.043 \\ 
10283026 &   58285.60 &  58286.00 &  $<$ 0.018 \\ 
10283027 &   58286.14 &  58286.81 &  $<$ 0.017 \\ 
10283028 &   58287.32 &  58287.32 &  $<$ 0.033 \\ 
10283029 &   58293.32 &  58294.05 &  $<$ 0.026 \\ 
10283030 &   58296.42 &  58296.97 &  $<$ 0.019 \\ 
10283031 &   58297.76 &  58297.89 &  $<$ 0.030 \\ 
10283032 &   58298.76 &  58299.96 &  $<$ 0.026 \\ 
10283034 &   58305.78 &  58307.00 &  $<$ 0.019 \\ 
10283035 &   58307.05 &  58307.92 &  $<$ 0.019 \\
\hline
All & 58010.35 & 58307.92  & 0.0072 $\pm$ 0.0012 \\
\hline
\end{tabular}
\end{center}
\caption{SWIFT Light curve for TRAPPIST-1}
\tablecomments{The data used to construct the light curve (plotted in Figure \ref{fig:lightcurve}) constructed from the SWIFT UVOT observations. Each observation is labeled by its SWIFT Observation ID number. Observation ID 10283022 present a detection of the UV flux, and all other observations present upper limits. Each observation ID denotes an observation taken continuously by SWIFT. The last row shows the result of stacking all 300 ks of observations, which yields a flux detection with a count rate of 0.0072 $\pm$ 0.0012 counts / sec.    }
\label{tab:swiftlc}
\end{table}

\subsubsection{SWIFT XRT Dataset}
\label{sec:xraylc}

\begin{figure*} 
   \centering
   \includegraphics[width=6.8in]{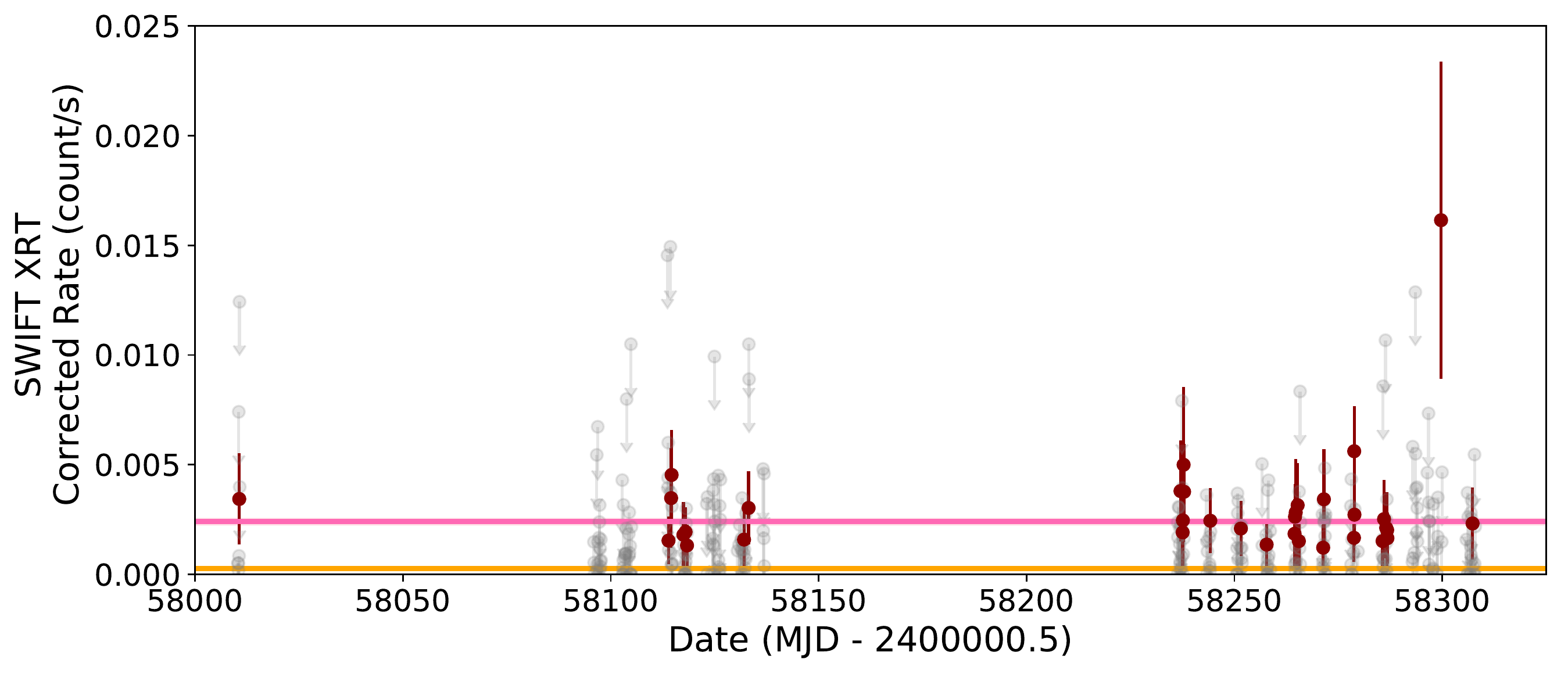}
   \caption{The X-ray light curve constructed using SWIFT XRT observations, in 5000 second bins. Detections are plotted as {dark red} points with error bars, and upper limits are plotted as grey points with error arrows extending downwards. {The detections correspond to flares, while each grey point corresponds to the dervied background level in that bin.} The average flux over the full 300ks of observations is plotted as an orange line, with the derived $1\sigma$ uncertainty plotted as a filled orange background region, {and the flux level during detections (red points) only is plotted as a pink line.} } \label{fig:xraylc}
\end{figure*}

In Figure \ref{fig:xraylc}, we present the detected photons {as a light curve. The data was collected in XRT's photon-counting mode.}
The light curve was produced by using a source extraction region centered at the sky position of TRAPPIST-1 and with a aperture radius of $\approx$45 arcseconds, and a nearby background region of $\approx$ 88 arcseconds. 
{After detected photons in the source and background regions were recorded, the final light curve was produced using the XSELECT program, by binning the source and the background light-curve to 5000 second bins and constructing the final corrected light-curve by subtracting the background off the source light-curve using XSELECT tool LCMATH.}

{The final result is plotted in Figure \ref{fig:xraylc}, and two colors of points are plotted: in red, bins when the source was detected, and in grey, bins where is was not. The detections correspond to flares, and the non-detections could correspond to quiescent periods or just periods where the background was too high to detect the flaring source.}

{The light-curve describes the flux level over time, but to find an averaged flux level, we do not fit the light-curve itself; instead, we stack the entire 300 ks of data and extract the flux from that. Excluding the flaring times and stacking all observations of periods of time corresponding to non-detections within each 5000 s bin (time periods plotted as grey upper limits), we find a baseline quiescent flux of 0.00026 counts/sec. Including only the detections (time periods plotted as red points), we find a detected flux of 0.0024 counts/sec. Using a stacked image of all 300 ks of data, we find an average flux of 0.00042 count/sec. These values are summarized in Table \ref{tab:starparams}. Importantly, the background is not large enough to account for the non-detections for our derived value of flaring and quiescent flux. }

\subsection{Derived Parameters}\label{sec:analysis}
From our light curve, we can derive several parameters describing the high-energy radiation environment of the system. As TRAPPIST-1 hosts at least seven planets whose habitability status and long-term atmospheric evolution depends tightly on their XUV insolation environments \citep{OMalley-James2017}, better constraints on these parameters allow better modeling of the evolution and history of the system.

\subsubsection{UV Flux}
\label{sec:uvw2flux}
Due to the large number of upper limits and low-significance detections, we measure the flux by utilizing a stacked image of the 300 ks of observations, centered on the sky position of TRAPPIST-1.
From the stacked image, we find a 5$\sigma$ detection of TRAPPIST-1, corresponding to a UVOT \texttt{uvw2} count rate of of 0.0072 $\pm$ 0.0012 CPS (counts/sec).
This count rate can be converted to a flux for a M-dwarf using the conversion rate given in Table 1 of \citet{Brown2016}, resulting in a measure of the flux at Earth in the UV of 8.4 $\times 10^{-15}$ erg/s/cm$^{2}$ (see Table \ref{tab:starparams}) {in the Swift \texttt{uvw2} waveband around $\sim1900$ Angstroms}. 

\subsubsection{X-ray Flux}
\label{sec:xrayflux}
The X-ray light curve has a large number of non-detections, and as such we also measure the X-ray flux by utilizing a stacked image of all photons received. 
From the stacked image, we find a background-corrected detection of TRAPPIST-1 with an XRT count rate of of 0.00042 $\pm$ 0.00005 CPS (counts/sec). {The values for the count rates during detections and non-detections are also given in Table \ref{tab:starparams}.}
This count rate can be converted to a flux for a M-dwarf using {the two-temperature model from \citep{Wheatley2017}. We do this by using a single conversion factor between Swift count rate and flux in erg/s/cm$^{2}$. This count rate was computed using XSPEC to recreate the model from \citep{Wheatley2017}, and then using the Swift gain scale and effective area of the telescope (which come from the redistribution matrix file and appropriate ancillary response files for Swift). In this case, we find that 1 count/s corresponds to 2.67e-11 erg/s/cm$^2$ for the Swift parameters and energy bandpass.}

{Using this conversion factor, we find that the average value of the X-ray flux over our measured full baseline was ($1.1 \pm 0.13$) $\times 10^{-14}$ erg/s/cm$^{2}$ (see Table \ref{tab:starparams}). The values during periods in which the source was detected and non detected are also given in Table \ref{tab:starparams}.}


\subsection{UV Flaring Activity}
A major goal of collecting a large quantity of \textit{Swift} UV data on TRAPPIST-1 was to study the star's flaring over time. 
Using the stacked image's derived flux as the baseline value, we find one event where the instantaneous measured flux value exceeds the baseline by more than $\sim3\sigma$. This occurs during observation ID 10283022, a roughly 19 ks observation taken starting on 2018-05-26 (see Table \ref{tab:swiftlc}).
All other observations yield either upper limits on the flux or measures consistent with the average flux level. 
{The integrated flux level across all our observations (and the upper limit on flux from roughly a quarter of our non-detections) is much lower than this solid detection.} Due to its significant observational baseline, our integrated 300 ks image is likely a good measure of the average UV flux emitted by the star, and we proceed with this averaged value in subsequent analysis in this paper. 

\subsubsection{EUV and XUV Flux}
EUV (100 - 900 Angstroms) observations of stars in general are made difficult by two factors: first, EUV flux is absorbed by the ISM mainly around 400 - 900 Angstroms; second, the Earth's geocoronal hydrogen is problematic when observing in the EUV.
As such, the EUV flux (the wavelength needed for studies of atmospheric mass loss, among other analyses) cannot generally be directly measured. 

However, the EUV flux can be extrapolated from wavebands where observations can be made using empirical scaling relations {\citep{SanzForcada2011, Linsky2014, SanzForcada2014, King2018}}. Since the sum of the EUV flux and X-ray flux controls the rate at which mass is lost from planetary atmospheres, {in this work we use our direct measure of the X-ray flux to derive extrapolated values for the EUV flux.}
{Our direct measure of the Swift UV flux (taken in a waveband centered near 2000 Angstroms) will ideally be useful for future photochemistry models \citep[see][]{Ranjan2017, Chen2019} but is not directly used in the XUV models that follow in this paper.}

From \citet{SanzForcada2011}, we have the empirically-derived mapping between EUV and X-ray luminosities:
\begin{equation}
    \log{L_{EUV}} = (4.80 \pm 1.99) + (0.860 \pm 0.073) \log{L_{X}},
    \label{eq:fluxscaling1}
\end{equation}
where this scaling relation was derived using \textit{GALEX} data. 
We also caution that the samples used to empirically derive Equation \ref{eq:fluxscaling1} included very few M-dwarfs and only a single late-type M-dwarf, so this extrapolation is a significant source of uncertainty in our analysis (as are many empirical scaling relations; \citealt{Drake2019}).

Using Equation \ref{eq:fluxscaling1}, we convert the x-ray flux computed in Section \ref{sec:xrayflux} to a EUV flux. The resultant final value of the combined XUV flux and the intermediate steps are presented in Table \ref{tab:starparams}.

\subsection{Comparison with Previous Work}
\citet{Wheatley2017} found that this star is a strong coronal X-ray source, with an X-ray flux of 2 or 4 $\times 10^{-14}$ erg s$^{-1}$cm$^{-2}$.




%

\citet{Peacock2018} created model spectra to derive model fluxes in the EUV, which cannot be observed directly, and found results varying between 1.32 - 17.4 $\times 10^{-14}$ erg /sec / cm$^2$, depending on the model used. From scaling relationships, previous work obtained values ranging between 0.7 - 5.8 $\times 10^{-14}$ erg /sec / cm$^2$ \citep{Bourrier2017, Wheatley2017}. 
Our observations imply a EUV flux at Earth of 2.6$
\times10^{-14}$ erg/s/cm$^{-2}$, which is roughly consistent with Model 1A of \citet{Peacock2018}. Model 1A was constructed by calibrating to the the Ly$\alpha$ observations for this star, while the other models were matched to \textit{GALEX} fluxes for M8 field stars. 
Our results are also consistent with the empirical model's prediction from \citet{Berger2010} (see Figure 3 of that paper, {which finds that for an M8 dwarf, $\log{L_{X}/L_{bol}}$ falls between -4 and -3.5, while we find $\log{L_{X}/L_{bol}} \approx$ -4}) for a star of this spectral type.

\subsubsection{Planetary mass loss rates}
Equation 1 of \citet{Bourrier2017} computes the mass loss rate {(see the original calculation in \citealt{Watson1981}, as well as \citealt{Erkaev2007,Lopez2012, Luger2015})} to have the form: 
\begin{equation}
    \dot{m_p} = \epsilon ( r_{XUV} / r_p )^{2} ( 3 F_{XUV} / 4 G \rho K )
    \label{eq:mdot}
\end{equation}
where $\epsilon$ is the heating efficiency (set to be equal to 0.1 in this work), $r_p$ is the true radius of the planet, $r_{XUV}$ is the effective planetary radius in the XUV waveband (which we assume is approximately equal to the true radius in this work, since the planet is rocky), $F_{XUV}$ is the XUV flux received by the planet, $G$ is the gravitational constant, and $K$ is a tidal enhancement factor {\citep{Erkaev2007}}, defined such that
\begin{equation}
    K = 1 - \frac{3}{2} \frac{r_{XUV}}{r_{roche}} +  \frac{1}{2}\frac{r_{XUV}^{3}}{r_{roche}^3}\,,
\end{equation}
where $r_{roche}$ is the planetary Roche Lobe (defined as $r_{roche}~=~(m_p / 2 M_{*})^{1/3} a$). For planets at short orbital periods, $K$ describes the enhancement in escape potential for gas in the planetary atmosphere: the proximity to the star and large effect of stellar gravity means that gas must reach the Roche radius to escape the planet. 

From Equation \ref{eq:mdot}, we can compute the expected mass loss rates for each of the planets in the TRAPPIST-1 system at the current day. These values are given in Table \ref{tab:sysparams}.
We note that although these rates are small at the current day, they are likely not constant over the age of the system. In the first 500 Myr of the system lifetime, the bolometric luminosity was much brighter \citep{Baraffe2015} and thus the mass loss rates equivalently higher (barring shielding against photo-evaporation due to the disk while the gas disk was still present).

\begin{table*}
\begin{center}
\begin{tabular}{ccc}
{Parameter} & {Value}  & {Source} \\
\hline
Mass, $M_{*}$  & 0.089 $\pm$ 0.006 $M_{\odot}$ & A \\
Radius, $R_{*}$  & 0.121 $\pm$ 0.003 $R_{\odot}$ & A \\
Luminosity, $L_{*}$ & (5.22 $\pm$ 0.17) $\times 10^{-4}$ $L_{\odot, bol}$& A \\
T$_{eff}$ & 2516 $\pm$ 41 K & A \\
P$_{rot}$ & 3.3 $\pm$ 0.14 days & B \\
Swift UVW2 & 0.0072 $\pm$ 0.0012  counts / sec & This Work, measured \\
Swift XRT  {(total)} & 0.00042 $\pm$ 0.00005  counts / sec& This Work, measured \\
{Swift XRT (detections)} & {0.0024 $\pm$ 0.00015}  counts / sec& {This Work, measured} \\
{Swift XRT (baseline)} & {0.00026 $\pm$ 0.00003}  counts / sec & {This Work, measured} \\
$F_{UV, uvw2, Earth}$  & (8.4 $\pm$ 1.4)\ $\times 10^{-15}$ erg s$^{-1}$ cm$^{-2}$ & This Work, measured \\ 
$L_{UV, uvw2}$ / L$_{*}$  & (7.4 $\pm$ 1.2) $\times 10^{-5}$ erg s$^{-1}$ cm$^{-2}$ & This Work, measured \\ 
$F_{X}$ (total)  & (1.1 $\pm$ 0.1)$\times 10^{-14}$ erg s$^{-1}$ cm$^{-2}$& This Work, measured \\ 
{$F_{X}$ (detections) } & (6.4 $\pm$ 0.4)$\times 10^{-14}$erg s$^{-1}$ cm$^{-2}$  & This Work, measured \\ 
{$F_{X}$ (baseline) } & (6.9 $\pm$ 0.8)$\times 10^{-15}$ erg s$^{-1}$ cm$^{-2}$ & This Work, measured \\ 
$L_{X}$ / L$_{*}$ (total)  & (1.0 $\pm$ 0.1) $\times 10^{-4}$  & This Work, measured \\ 
{$L_{X}$ / L$_{*}$ (detections) } & (5.6 $\pm$ 0.4)$\times 10^{-4}$  & This Work, measured \\ 
{$L_{X}$ / L$_{*}$ (baseline) } & (6.1 $\pm$ 0.7)$\times 10^{-5}$ & This Work, measured \\ 
{$L_{EUV}$ / L$_{*}$ (total)}  & (2.4 $\pm$ 0.3)$\times 10^{-4}$  & This Work, extrapolated \\ 
{$L_{XUV}$ / L$_{*}$ (total)} & (3.4 $\pm$ 0.4)$\times 10^{-4}$  & This Work, extrapolated 
\end{tabular}
\end{center}
\caption{Stellar Parameters for TRAPPIST-1}
\tablecomments{The stellar parameters for planet host TRAPPIST-1, a late M-dwarf. {In this work, the bolometric luminosity of TRAPPIST-1 is defined by the value of $L_{*}$ given in the table \citep{VanGrootel2018}.} A: Values come from \citet{VanGrootel2018}. B: Value from \citet{Luger2017}. }
\label{tab:starparams}
\end{table*}

\section{Theoretical Implications}\label{sec:planets}
There has been a significant amount of follow-up observations on the TRAPPIST-1 system, resulting in high-precision measurements of the planetary orbital parameters. The earliest analyses of the system, in which the planet discoveries were announced, reported only upper limits for eccentricities \citep{Gillon2017}, but recent work using a longer observational baseline finds the eccentricities of all planets are non-zero, to significance levels varying between 2$\sigma$ and 10$\sigma$ \citep{Grimm2018}.

TRAPPIST-1 b's eccentricity being 2$\sigma$ inconsistent with zero is particularly intriguing, since its extremely short current-day orbital period indicates that it will be susceptible to tidal decay of eccentricity and semi-major axis. 
Many papers \citep[i.e][]{Papaloizou2018} have conducted their analysis of the TRAPPIST-1 system by assuming null eccentricities for all planets, as such a configuration boosts dynamical stability and increases the timescale on which the system can be studied numerically without becoming destabilized. 
In this work, however, we consider the implication of the potential non-zero eccentricity on the tidal parameters of the planets, and discuss the {impact these arguments have on our understanding of the planetary eccentricities}.

\subsection{Analytic Modeling of Tidal and Atmospheric Evolution}
\label{sec:analysis_secular}
The rate of orbital radius change expected due to tides has two components: one due to the tides raised on the planet, and another due to the tides raised on the host star. 
This change may manifest as either a decay \citep[i.e.,][]{Barnes2008, Silburt2015} or an expansion of the orbits \citep[as for the moon;][]{md99}, depending on the specific tidal, spin, and orbital parameters of the involved bodies. 
In this section, we consider whether the currently observed planetary eccentricities in the TRAPPIST-1 system are likely to persist on timescales of order the age of the system by using TRAPPIST-1 b as the boundary condition. If TRAPPIST-1 b, the most susceptible planet to eccentricity damping, can maintain its eccentricity under the effects of both tidal evolution due to interactions with its host star and atmospheric mass loss due to stellar EUV insolation, then likely all planets can.

The magnitude of these effects can be estimated using the following expression from \citet[][]{Goldreich1963} (with expressions written in a functionally identical form to that used in \citealt{Barnes2008}):
\begin{equation}
\begin{split}
da/dt = \Big( &-\frac{63}{2}\frac{1}{Q' m_p}  \sqrt{GM^3} r_p^5 e^2 + \\
 & \frac{9}{2}\frac{\sigma}{Q_{*}'}  \sqrt{G/M} R^5 m_{p} \Big) a^{-11/2},
\end{split}
\label{eq:dadt}
\end{equation}
where $Q' = 3Q / 2k_{2}$ is three halves times the ratio of the effectiveness of a body's rate of energy dissipation due to tidal distortions ($Q$) to the the love number ($k_{2}$, describing a body's central concentration), $m_p$ denotes a planetary mass, $r_p$ denotes a planetary radius, $G$ is the gravitational constant, $M$ denotes the stellar mass, $R$ the stellar radius, $a$ denotes the planetary semi-major axis, $e$  the planetary eccentricity, and the term $\sigma = sign(2\Omega - 3n)$ (where $\Omega$ is the stellar spin rate and $n$ the planetary mean motion) describes the relative frequencies of the stellar rotational period and planetary orbital period. 
In \citet{Gillon2016}, the TRAPPIST-1 rotation rate was estimated using ground-based data to be $P_{rot}= 1.40$ days, but follow-up work using data from the K2 spacecraft \citep{Luger2017, Vida2017} provided an updated value of $P_{rot}= 3.3$ days, which we use in this analysis. 
This value of $P_{rot}= 3.3$ days (giving the star a rotational period situated in between the orbital periods of planets c and d) means that the sign term $\sigma$ in Equation \ref{eq:dadt} cannot be neglected, as different planets will be affected differently by tides raised on the star. We do note that as $Q_{*}$' is very large ({in this work, we use $Q_{*}$' $ = 10^5$, while} estimates for stars range between $10^{4}$ and $10^{8}$; \citealt{Jackson2009}), the second term of this equation will have a negligible effect compared to the first term, leading to tides raised on the planet rather than those raised on the star having the dominant effect in the orbital evolution of the planet. 

A secondary paired equation can be written to describe the evolution of eccentricity over time \citep{Goldreich1963},
\begin{equation}
\begin{split}
de/dt = \Big( &-\frac{63}{4}\frac{1}{Q' m_p}  \sqrt{GM^3} r_p^5 + \\
 & \frac{171}{16}\frac{\sigma}{Q_{*}'}  \sqrt{G/M} R^5 m_{p} \Big) e\ a^{-13/2},
\end{split}
\end{equation}
where the first term is due to tides raised on the planet and the second term is due to tides raised on the star. 
Combined with Equation \ref{eq:dadt}, these two expressions describe the decay of a planet's orbit, which will occur more quickly for planets which dissipate their stored energy very quickly (i.e., more rocky planets, or those with lower $Q'$).

All of the parameters in both of these equations can be measured from the TRAPPIST-1 light curve (radii directly, and masses via transit timing variations as in \citealt{Grimm2018}, and stellar parameters using the K2 data as in \citealt{Luger2017}) except for the $Q'$ ratio, which describes the physical structure of the planet. 
Since the change in orbital elements scales roughly linearly with $Q$', slightly different interior structures can result in different timescales for the damping of orbital eccentricity.
Then, based on the currently measured orbital parameters and an estimate of the age of the system, the true value of the $Q'$ ratio can be estimated on the basis of a dynamical analysis (as is commonly done for hot Jupiters, whose $Q'$ values must allows them to survive through the age of the system until our current observations; \citealt{Barker2009}). In the case of TRAPPIST-1, the relatively short orbital periods, combined with the non-zero eccentricities (if confirmed), would yield strong indirect constraints on the planets' internal structure. 

There is an added complication on the observed dynamics due to the evaporating atmosphere. 
If the XUV flux is high enough that the planets are losing atmospheric mass, then our current-day estimates of the orbital decay might be overestimated (if only slightly, given that the computed mass loss rates are very slow). 
To fully encapsulate the tidal dynamics of this system, all the factors enumerated above must be considered, which we do by solving the evolution as a coupled set of differential equations, as follows.

To model the expected evolution of eccentricity and semi-major axis for TRAPPIST-1 b (for now, neglecting planet-planet interactions and treating the short-period TRAPPIST-1 b as the edge case where constraints will be most easily derived), we combine and solve the system of differential equations including the damping of semi-major axis and eccentricity due to tides, as well as the atmospheric loss rate computed in Section \ref{sec:data}: 
\begin{equation}
\begin{cases}
de/dt = \Big( \frac{-63}{4 Q' m_p}  \sqrt{GM^3} r_p^5 +  \frac{171}{16}\frac{\sigma \sqrt{G}}{Q_{*}' \sqrt{M}}  R^5 m_{p} \Big) e\ a^{-13/2}\\
da/dt =  \Big( \frac{-63}{2 Q' m_p}  \sqrt{GM^3} r_p^5 e^2 + \frac{9}{2}\frac{\sigma\sqrt{G}}{Q_{*}'\sqrt{M}}   R^5 m_{p} \Big) a^{-11/2} \\
 dm_p/dt = \epsilon ( r_{XUV} / r_p )^{2} ( 3 F_{XUV} / 4 G \rho K ) \\
\end{cases}
\label{eq:simult}
\end{equation}
where for now we assume $F_{XUV}$ is constant over time (this is a good approximation when modeling evolution in M-dwarfs taking place when the star is more than 1 Gyr old; see the models of {\citealt{Baraffe2015}}). 
We note that for larger mass stars where the luminosity does change appreciably over the main sequence lifetime of the star, an additional coupled equation $dL_{*}/dt$ should be added to the system, where the stellar luminosity can be numerically modeled (using, for example, the models of \citealt{Baraffe2015}). {Note that this was done in \citet{Luger2015b, Luger2015,Lincowski2018}.}

To study the pre-main sequence behavior of these planets, this variable luminosity would need to be accounted for (see Section \ref{sec:vplanet}). 
We also assume that the TRAPPIST-1 planets, being similar in radius to the Earth, are terrestrial and will lose mass from their atmospheres, leading to the planetary $Q$' ratios being constant over the age of the system. 

\begin{table*}
\begin{center}
\begin{tabular}{cccccccccccc}
\textbf{Planet} & \textbf{Period} & \textbf{Radius} & \textbf{Mass} & \textbf{e}  & \textbf{XUV Flux}  & \textbf{Mass loss}  & \textbf{Mass loss}  \\
 & (days) & (R$_{\oplus}$) & (M$_{\oplus}$) & &(erg/cm$^2$/s) & (gram/sec) & ($M_\oplus$ /Gyr) \\
\hline
TRAPPIST-1 b &      1.510876 &           1.121 &         1.017 &  0.00622 &             2935.128486 &    8.286578e+08 &               0.004376 \\ TRAPPIST-1 c &      2.421807 &           1.095 &         1.156 &  0.00654 &             1564.865012 &    3.622544e+08 &               0.001913 \\ TRAPPIST-1 d &      4.049959 &           0.784 &         0.297 &  0.00837 &              788.454573 &    2.607478e+08 &               0.001377 \\ TRAPPIST-1 e &      6.099043 &           0.910 &         0.772 &  0.00510 &              456.452067 &    9.081431e+07 &               0.000480 \\ TRAPPIST-1 f &      9.205585 &           1.046 &         0.934 &  0.01007 &              263.598217 &    6.583289e+07 &               0.000348 \\ TRAPPIST-1 g &     12.354473 &           1.148 &         1.148 &  0.00208 &              178.116325 &    4.784535e+07 &               0.000253 \\ TRAPPIST-1 h &     18.767953 &           0.773 &         0.331 &  0.00567 &              102.035492 &    2.902111e+07 &               0.000153 
\end{tabular}
\end{center}
\tablecomments{Planetary parameters for the planets in the TRAPPIST-1 system. Orbital periods, planetary masses, radii, and eccentricities come from \citet{Grimm2018}; errors from that paper are not reproduced in this table for brevity. The XUV flux incident on each planet and the resultant planetary atmospheric mass loss rates were derived in Section \ref{sec:data} of this work. } 
\label{tab:sysparams}
\end{table*}

\begin{figure} 
   \centering
   \includegraphics[width=3.4in]{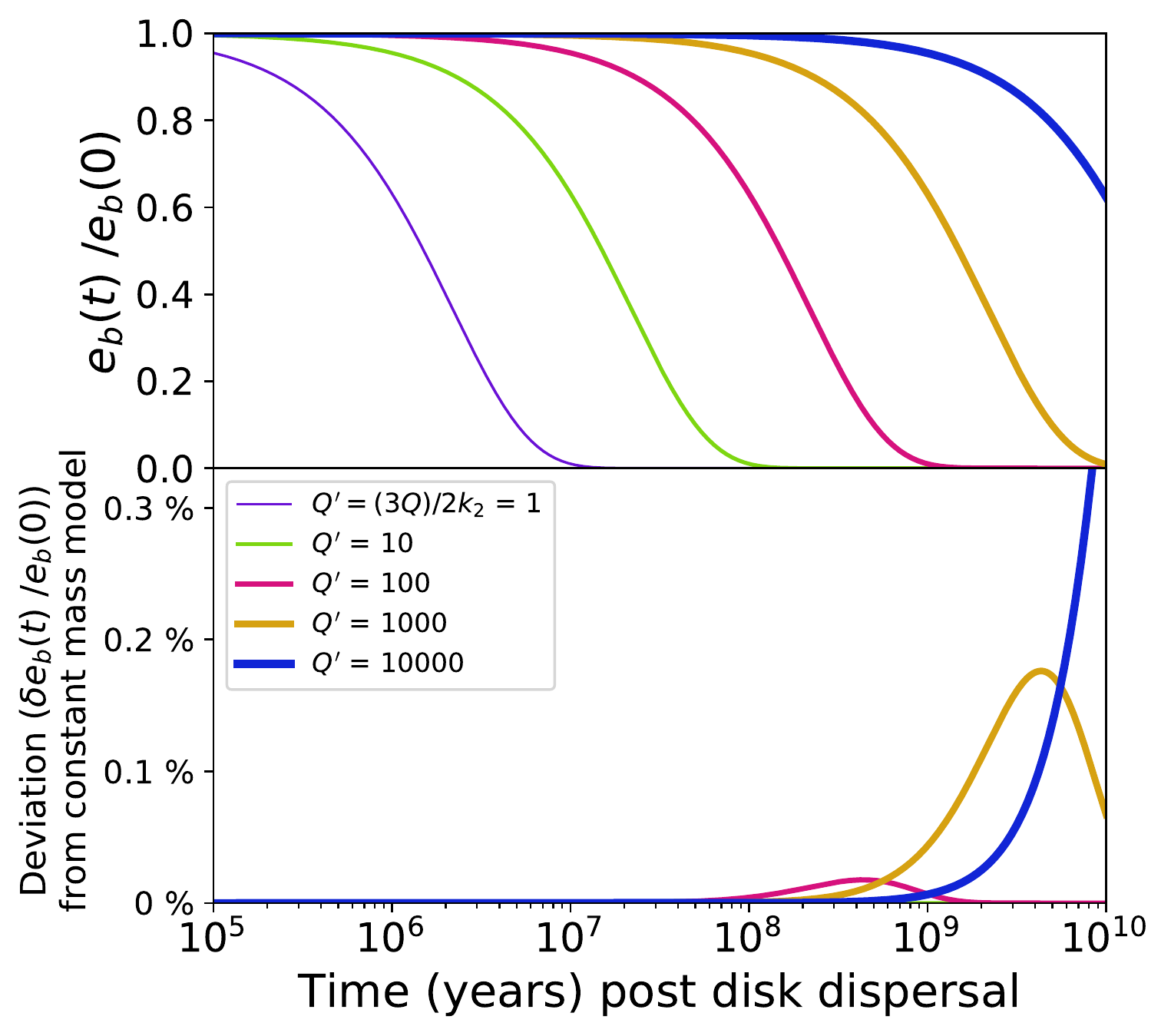}
   \caption{{(Top panel) The fractional eccentricity dissipation over time for TRAPPIST-1 b, compared to the initial value at the time of disk dispersal $e_{b}(0)$. The evolution was computed by simultaneously solving the system given in Equation \ref{eq:simult} with planetary semi-major axis fixed to its currently measured value. The forward evolution demonstrates the expected damping in eccentricity expected for this short-period planet by tidal parameter $Q'$. (Bottom panel) The percentage deviation between our model (which includes the mass loss, as planetary masses decrease at a rate given by Equation \ref{eq:mdot}) and our model with the mass loss rate set to zero. When mass loss is accounted for, the time at which the planet circularizes occurs earlier than if it is ignored, and the difference is larger for larger values of $Q'$.}}
   \label{fig:ecc_decay}
\end{figure}

{In Figure \ref{fig:ecc_decay}, we show the evolution of eccentricity of TRAPPIST-1 b over time under two major assumptions: first, the evolution is modeled starting at the time of disk dispersal, and we assume that the entire system of planets had attained roughly their current orbits at that time; second, we assume that the XUV luminosity of TRAPPIST-1 is constant over this time (and thus the evolution begins after the star has reached the main sequence). This analysis (and those that follow in this section) are intended to model the dynamical evolution of the system after the planets reach their final orbital periods, and determine how the tidal parameters $Q'$ of each of the planets affect that evolution. Figure \ref{fig:ecc_decay} shows that the fractional eccentricity of TRAPPIST-1 b is damped fairly quickly (i.e., for a non-zero eccentricity to be maintained, another process must be at play). }

\begin{figure} 
   \centering
   \includegraphics[width=3.4in]{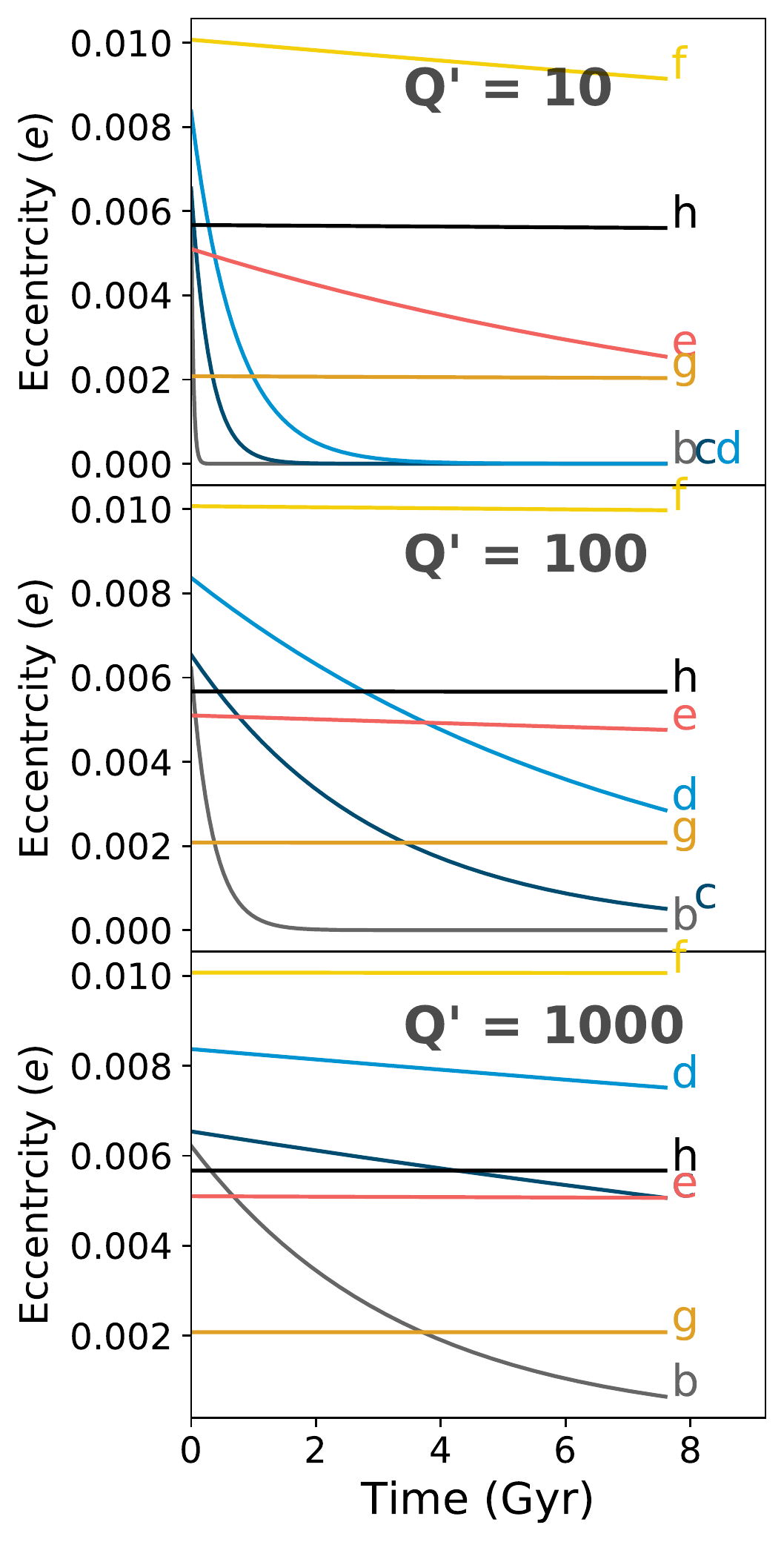}
   \caption{The eccentricity evolution over time for each of the planets in the TRAPPIST-1 system, with the planet name labeled at the end of its evolution (after 7.6 Gyr). These curves were computed assuming that the planets formed \emph{in situ}. The three panels denote three possible {values for the ratio $Q'$}. For smaller values of $Q'$, the eccentricity of the inner planets damps more quickly, as expected.  }
   \label{fig:noint_allplanets}
\end{figure}

Additionally, altering the initial parameters of this model will not change the exponential damping time scale, meaning that even if the eccentricities had been larger initially, they should have damped to zero by the current day {(that is, the fractional evolution shown in Figure \ref{fig:ecc_decay} will hold for a range of starting planetary eccentricities, which does not necessitate that the planets had any particular values when the system was assembled.) The bottom panel of Figure \ref{fig:ecc_decay} shows the percentage difference when the changing planetary masses are and are not considered; although the deviations between the two models are very small, the magnitude of the deviations are a function of planetary $Q'$, meaning that mass loss must be modeled when attempting to constrain $Q'$. In systems with higher mass loss rates, this effect would become even more important.  } 

TRAPPIST-1 b resides at current day on a very short orbital period (1.5 days, which corresponds to a semi-major axis of $\sim0.011$ AU).
At the same time, the \citet{Grimm2018} measurement of the current-day eccentricity of TRAPPIST-1 b is 0.00622 $\pm$ 0.00304. Among the planets in the TRAPPIST-1 system, planet b (having a non-zero eccentricity and the shortest orbital period) will be the most susceptible to eccentricity damping due to interactions with its host star. 

In Figure \ref{fig:noint_allplanets}, we plot the evolution of eccentricity for each planet as it depends on the $Q'$ ratio chosen for that particular planet. The timescales upon which the eccentricity will be damped can be determined from Equation \ref{eq:simult} and the chosen value of $Q'$.

\begin{figure} 
   \centering
   \includegraphics[width=3.4in]{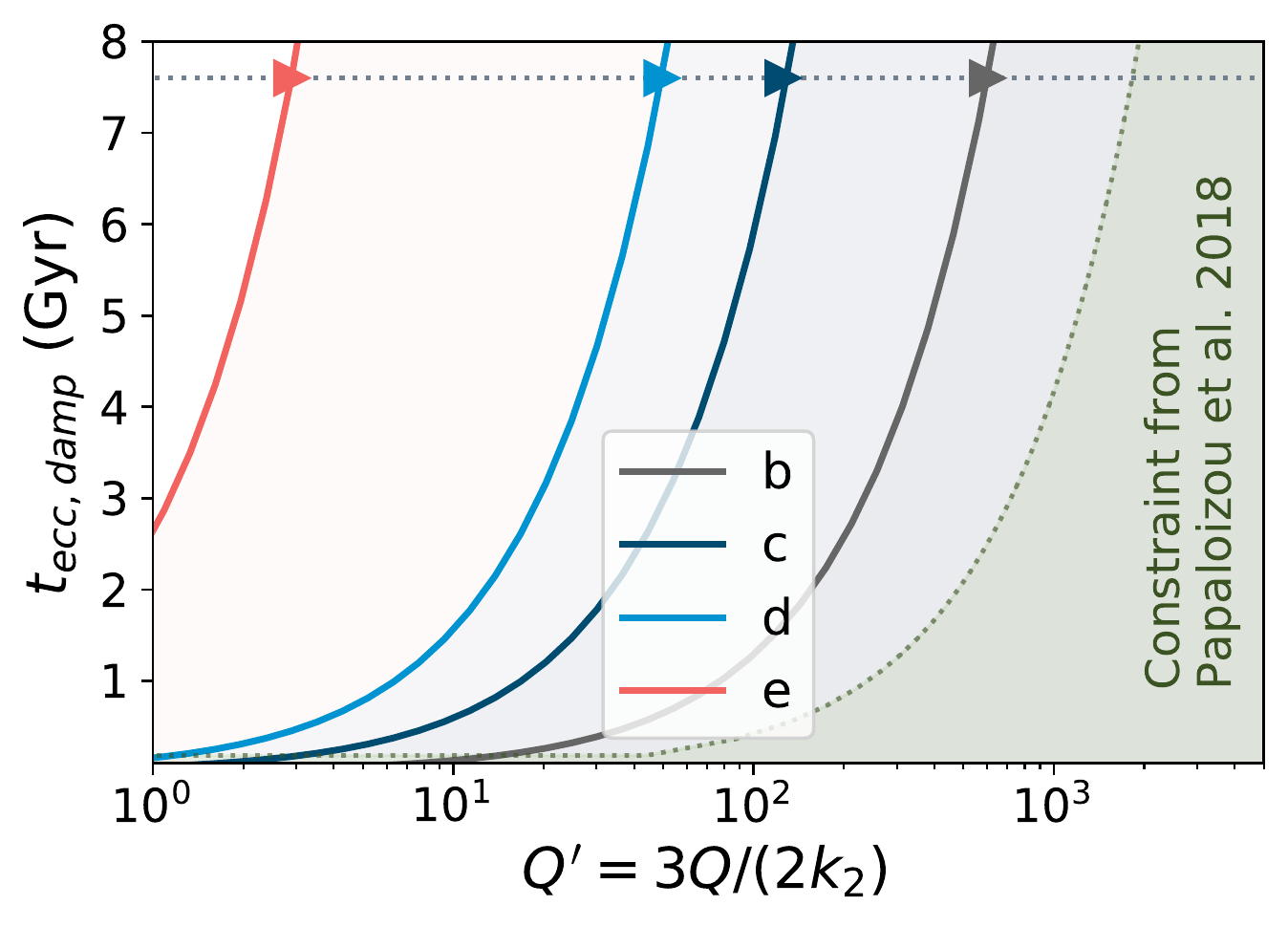}
   \caption{The allowable lower limit of $Q'$ for each planet in the TRAPPIST-1 system for the planets for which such constraints can be made, determined by computing the timescale within which the planetary eccentricity will damp to values which cannot be differentiated from zero with current fit precision.   }
   \label{fig:not_best_plot}
\end{figure}

In Figure \ref{fig:not_best_plot}, we plot the dependence of the final system eccentricity at various times as it depends on the planetary $k/Q$ ratio.
We also plot for comparison the range of allowable $Q'$ derived in \citet{Papaloizou2018}. In that work, authors derived the values of $Q'$ which allowed the planets to keep semi-major axis values close to their current resonant values over the course of the migration; their derived limit of $Q' >1.2 \times 10^{3} (t\ /\ 5\ \rm Gyr)$ reduces to $Q' > 1.8 \times 10^{3}$ for system age $t = 7.6$ Gyr \citep{Burgasser2017}. 
In this work, we plot the values of $Q'$ that allow the eccentricities to be observationally determined to be non-zero, disallowing constraints where the departure from zero eccentricity could not be measured with the precision of the current TTV fits of \citet{Grimm2018}. 
This analysis presents a lower limit for $Q'$ for the inner four planets (constraints cannot be made for the outer three planets, as any $Q'$ allows the observed eccentricity values). 

\subsection{Numerical Modeling of Tidal and Atmospheric Evolution}

\label{sec:vplanet}
In the previous section, we considered the evolution of the system using a simple model which can be easily computed and used in conjunction with Monte Carlo techniques to robustly explore the (in this case, $k/Q$) relevant parameter spaces in the system. In this section, we turn to numerical techniques to model how our derived value of the XUV luminosity of the host star TRAPPIST-1 affects the capabilities of its seven planets to maintain surface water, and how the $Q'$ ratio affects ocean retention. 

We use the Virtual Planet Simulator \texttt{VPLanet} \footnote{\url{https://github.com/VirtualPlanetaryLaboratory/vplanet}} \citep{Barnes2019} to numerically consider the evolution of the TRAPPIST-1 system in the presence of tides and atmospheric mass loss, the same processes considered in the previous section. 
In addition to the processes described in the previous section, \texttt{VPLanet} can simultaneously model the stellar, orbital, rotational, and even climate evolution of the system over time \citep{Barnes2019}, and has been used to study magma oceans on the TRAPPIST-1 planets (Barth et al, in prep). Here, we use \texttt{VPLanet} modules \texttt{EqTide} (which uses the same constant-phase-lag model used in the analytic modeling above to model tidal evolution, but using the full set of equations rather than just those for $da/dt$ and $de/dt$; see Appendix E of \citealt{Barnes2019})) and \texttt{AtmEsc} (which computes the energy-limited and diffusion-limited atmospheric escape; see Appendix A of \citealt{Barnes2019}). Between these two modules, we are able to self-consistently model the oceanic, atmospheric and tidal evolution of the (decoupled) planets.

We use as input to \texttt{VPLanet} the {saturated} $L_{XUV} /L_{*}$ ratio. 
{The ratio we computed in Section \ref{sec:uvw2flux} is the present-day value of this ratio; however, this ratio is known to decrease once the star spins down and leaves the `saturated' regime, after which $L_X/L_{bol}$ decays exponentially \citep{Pizzolato2003} - note that this decay is of the ratio $L_X/L_{bol}$ and is separate from any global change in $L_{bol}$. In \citet{Fleming2019}, the authors use the \citet{Wheatley2017} value of $L_{XUV}$ luminosity and find that TRAPPIST-1 is likely (at a $\sim$40\% probability) still in its saturated phase. The \citet{Wheatley2017} measurement is a factor of two larger than the value computed in this work. To find the saturated XUV ratio which corresponds to the present-day XUV ratio we measure in this work, we use the best-fit values from \citet{Fleming2019} of the saturation timescale $t_{sat} = 6.85$ Gyr, $\beta_{sat}$ = -1.16 (when $\beta$ is an exponent controlling how $L_{XUV}$ decays after saturation ends; see Equation 1 of \citealt{Fleming2019}).}

{Using these best-fit values and our own present day measurement of $L_{XUV}$ / L$_{*}$ = (3.4 $\pm$ 0.4)$\times 10^{-4}$, we model the evolution of the star TRAPPIST-1 alone (no planets) and find that our value (combined with the best-fit model from \citealt{Fleming2019}) corresponds to a saturated $L_{XUV}$ / L$_{*}$ ratio of $L_{XUV}$ / L$_{*}$ = 3.7 $\times 10^{-4}$. We use this value as the input to VPlanet, which combined with the aforementioned \citealt{Fleming2019} parameters for $t_{sat}$ and $\beta$ causes the VPlanet luminosity evolution to reproduce our observed value after 7.6 Gyr. }
We also use as inputs the currently measured best-fit planet parameters (taken from \citealt{Grimm2018}). 
\texttt{VPLanet} simultaneously solves the coupled ordinary differential equations describing the tidal and atmospheric evolution of the planets. We note that in this work, we do not solve for the full N-body evolution of the planets. We begin the planets with 10 Earth oceans worth of surface water (as compared to \citealt{Lincowski2018}, in which authors conducted a similar analysis of and started the planets with 20 Earth oceans of water). 

\begin{figure} 
   \centering
   \includegraphics[width=3.4in]{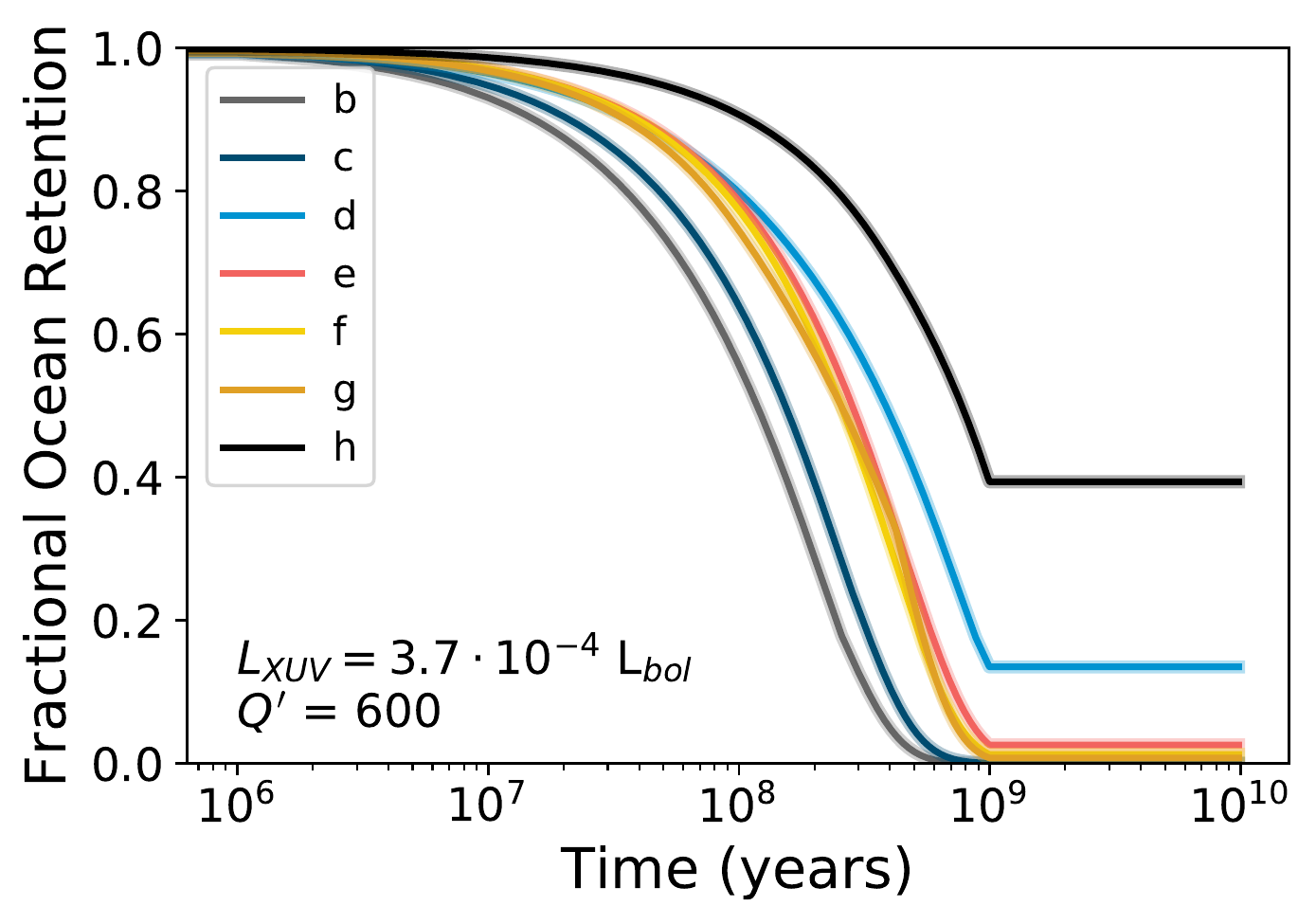}
   \caption{For the theoretically allowable value $Q' = 600$, the \texttt{VPLanet} evolution of the surface water composition over the first 5 Gyr of the TRAPPIST-1's system evolution. {For this combination of $Q'$ and $L_{XUV}$ / L$_{*}$, substantial oceans could be retained on planets d and h, and non-zero oceans retained on e, f, and g, while b and c will have totally lost their oceans. Some UV-motivated} estimates of the current-day habitable zone of the system suggest that liquid water should only exist on planets e and beyond \citep{OMalley-James2017}, {while others \citep{Barr2018} find that d may also retain an ocean, consistent with our results}.}
   \label{fig:waterloss1}
\end{figure}

Following \citet{Papaloizou2018}, we set the $Q'$ values to be the same for all seven planets in each integration. We find that varying the values of $Q'$ (within our previously derived allowable range) does not result in appreciably different ocean loss rates; the results of this numerical evolution, presented in Figure \ref{fig:waterloss1}, were initialized with a $Q' = 600$ for all planets.
We note that with $Q'$ values below our allowed range (for example, $Q' = 60$) the outcome resolved by \texttt{VPLanet} is collision with the central star. \ref{sec:analysis_secular} to be more feasible given the current day planetary eccentricities.
We see that with our extrapolated value of the XUV flux, the three inner planets do not retain surface water, losing it instead during the first Gyr of TRAPPIST-1's lifetime. 
We do note that ground water could potentially replenish lost surface water, and the above analysis does not consider this potential effect. 

%
%
%
%
%
%
%
%
%
%
%
%
%
%
%
%
%
\section{Discussion}
\label{sec:discuss}

In this work, we report on a continuum UV flux measurement of TRAPPIST-1, based on 300 ks of \textit{Swift} UVOT data collected over a baseline of approximately one year (during which we isolate one possible flaring event). The measured value, of (8.4 $\pm 1.4$) $10^{-15}$ erg/cm$^{2}$/sec {in the Swift \texttt{uvw2} waveband centered around $\sim1900$ Angstroms}, is consistent with previous estimates for the specific star TRAPPIST-1 \citep{Wheatley2017, Bourrier2017}.
We also present an X-ray flux measured with \textit{Swift} XRT data of $(1.1 \pm 0.13) \times 10^{-14}$ erg/cm$^{2}$/sec, which corresponds to a lower $L_{X}/L_{bol}$ ratio than found in previous work \citep{Wheatley2017}. {Our derived values for the current EUV and XUV luminosities for TRAPPIST-1 as a fraction of the total luminosity are $(2.3 \pm 0.3) \times 10^{-4}$ and $(3.4 \pm 0.4) \times 10^{-4}$, respectively.}

\subsection{Conclusions on Planetary Eccentricities}

As illustrated in Figure \ref{fig:ecc_decay}, the timescale over which the exponential damping of eccentricities occurs depends on the ratio $Q' = 3Q / 2 k_{2}$ of the planet. 
The planetary eccentricities measured in \citet{Grimm2018} were found to be non-zero by a significant margin. Under the assumption that these reported values are valid, we can make the following statements regarding the planetary eccentricities. One of the following two options must hold: 

\begin{itemize}
    \item The planets (particularly the inner planet, b) have a significant envelope capable of storing energy (and hence a larger ratio $Q' = 3Q / 2 k_{2}$).
    \item The planetary eccentricities have already been damped and the non-zero eccentricities measured in \citet{Grimm2018} are not physical, but rather a result of the limited amount of data available at the present time. 
\end{itemize}%
The first option is supported by the conclusions of \citet{Papaloizou2018} {and also \citet{Brasser2019}}, who also find that the $Q'$ values for the TRAPPIST-1 planets must be large in order for the resonant chain to persist.

The planetary eccentricities considered above arise due to planet-planet interactions. Our dynamical analysis does not include planet-planet coupling, although the resonant state of the system suggests that the planets are indeed coupled. We can see from the numerical simulations of \citet{Papaloizou2018} that the net eccentricity will damp over time. At the same time, eccentricities of the current magnitudes appear to cause the system to go unstable over relatively short ($<$1 Gyr) timescales \citep{Tamayo2017a}.

\subsection{Conclusions on Planetary Tidal Qualities}

Assuming that the planetary eccentricities are physical, their current values allows us to draw constraints on the $Q'$ parameter for each planet. 

The age of the star is estimated to be 7.6 $\pm$ 2.2 Gyr \citep{Burgasser2017}. Figure \ref{fig:not_best_plot} illustrates how the constraints on $Q'$ depend on the age of the system: For a system younger than 7.6 Gyr, we get a smaller (less strict) lower limit for the allowable $Q'$ ratio, as the eccentricity would have needed to be maintained for a smaller amount of time.  

For TRAPPIST-1 b, we get the strongest constraint: for a 7.6 Gyr system age, we find that $Q'_{b} > 600$. For a typical $k_{2} = 0.3$, this result corresponds to $Q_{b} > 120$. For planets c, d, and e, the limits are less strict. In our own solar system, the terrestrial planets have $Q$ values estimated to be $Q~<~190$ for Mercury, $Q~<~17$ for Venus, $Q~\approx~12$ for Earth, and $Q~<~26$ for Mars \citep{Goldreich1966}. In this context, our limit for TRAPPIST-1 b suggests a {planet with Mercury-like dynamical properties and reactions to tidal deformation rather than one with properties more similar to Earth or Mars}. The constraints on planets c, d, and e do not exclude Earth-like dynamical properties such as concentration and tidal dissipation susceptibilities. 

\citet{Papaloizou2018} find that in order for the resonances to persist on Gyr time scales while the planets migrate, the composite parameter $3 Q / (2 k_2)$ must be greater than $10^2 - 10^3$.  This work uses separate considerations --- allowing the measured eccentricities to persist (which does not depend on migration timescales). Nonetheless, we find that the limit for TRAPPIST-1 b is similar to that found in \citet{Papaloizou2018}, where Figure \ref{fig:not_best_plot} provides a direct comparison. On the other hand, our limits for the outer planets (c, d, and e) are significantly less stringent. Moreover, our constraints are agnostic regarding the assembly mechanism of the system, relying only on the existence of eccentricity damping after the disk has dissipated.

\section{Summary}\label{sec:sum}

This paper develops a combined analytical treatment of the tidal evolution and XUV-driven atmospheric mass loss for close-in exoplanets, and applies it to the benchmark extrasolar planetary system TRAPPIST-1. We first estimate the average UV flux level of the TRAPPIST-1 star to be \uvflux\ (observed at Earth). {Then, we measure the Swift X-ray flux of TRAPPIST-1, which corresponds to an extrapolated XUV luminosity fraction of \xuvfrac.} With this derived value, we calculate the atmospheric mass loss rates for the seven known planets in the TRAPPIST-1 system, and estimate their ranges of allowed $Q' = 3 Q / 2 k_{2}$ ratios, under the assumption that the currently measured orbital eccentricities from \citet{Grimm2018} are correct. For the outer three planets, no meaningful constraints can be made, and we cannot exclude Earth-like tidal $Q$ values for planets c, d, and e. For TRAPPIST-1 b, we find a lower limit of $Q'~>~600$, a limit larger that the $Q'$ values of our own solar system's terrestrial planets. Further, our \texttt{VPLanet} analysis demonstrates that our measured UV and X-ray luminosities suggest that the third planet, TRAPPIST-1 d, may retain a surface ocean.  The analytical method developed in this paper can be extended to planets with higher mass loss rates in order to study their coupled atmospheric and orbital evolution. 

\medskip

This project used the following software: 
pandas \citep{mckinney-proc-scipy-2010}, IPython \citep{PER-GRA:2007}, matplotlib \citep{Hunter:2007}, scipy \citep{scipy}, numpy \citep{oliphant-2006-guide}, Jupyter \citep{Kluyver:2016aa}, 
vplanet \citep{Barnes2019}, ds9 \citep{ds9}

{Code, data, and supplemental figures related to this work can be found at \url{https://github.com/jxcbecker/trappist1}.} 

\acknowledgments
We thank the referee for a useful report. We thank Sarah Peacock and Francesco Haardt for assistance. We also thank Andrew Vanderburg, Eric Agol, {Konstantin Batygin, Erik Petigura,} Erin May, and Beate Stelzer for useful conversations. During this work, J.C.B.~has been supported by the Leinweber Center for Theoretical Physical Graduate Fellowship and the Heising-Simons \textit{51 Pegasi b} postdoctoral fellowship. RB acknowledges support from the NASA Astrobiology Program Grant Number 80NSSC18K0829.



\bibliographystyle{apj}

\end{document}